\documentclass[aip,apl,reprint,superscriptaddress,showpacs]{revtex4-1}

\usepackage{graphicx}
\usepackage{ulem}
\usepackage{color}

\begin{document}

\title{Enhancing high-temperature thermoelectric properties of PtAs$_2$ by Rh doping}  

\author{Kazutaka Kudo} \email{kudo@science.okayama-u.ac.jp}
\author{Seiya Nakano}
\author{Tasuku Mizukami}
\affiliation{Department of Physics, Okayama University, Okayama 700-8530, Japan}
\affiliation{Advanced Low Carbon Technology Research and Development Program (ALCA), Japan Science and Technology Agency (JST),
Tokyo 102-0076, Japan}
\author{Toshiro Takabatake}
\affiliation{Department of Quantum Matter, ADSM, Hiroshima University, Higashi-Hiroshima 739-8530, Japan }

\author{Minoru Nohara} \email{nohara@science.okayama-u.ac.jp}
\affiliation{Department of Physics, Okayama University, Okayama 700-8530, Japan}
\affiliation{Advanced Low Carbon Technology Research and Development Program (ALCA), Japan Science and Technology Agency (JST),
Tokyo 102-0076, Japan}

\date{\today}

\begin{abstract}
The effects of Rh doping on the thermoelectric properties of Pt$_{1-x}$Rh$_x$As$_2$ ($x$ = 0, 0.005, and 0.01) with pyrite structure were studied by conducting measurements of electrical resistivity $\rho$, Seebeck coefficient $S$, and thermal conductivity $\kappa$.
The sample with $x$ = 0.005 exhibited large $S$ and low  $\rho$, resulting in a maximum power factor ($S^2/\rho$) of 65 $\mu$W/cmK$^2$ at 440 K.
The peculiarly shaped ``corrugated flat band'' predicted for PtSb$_2$ might explain the enhanced thermoelectric properties of doped PtAs$_2$.
\end{abstract}


\maketitle

Thermoelectric power generation has attracted much recent attention as a means for directly converting waste heat into electricity; however, its successful commercial application will require the development of thermoelectric materials with high efficiency.
Thermoelectric efficiency can be evaluated using the dimensionless figure of merit $ZT = S^2T/\rho\kappa$, where $S$ is the Seebeck coefficient, $\rho$ is electrical resistivity, $\kappa$ is thermal conductivity, and $T$ is absolute temperature.
In this formula, the power factor (PF) = $S^2/\rho$ expresses the upper limit on the amount of electrical power that can be drawn from a material; to efficiently generate thermoelectricity, therefore, it is important to have both a large Seebeck coefficient $S$ and a low (metallic) electrical resistivity $\rho$.
This is not easy to obtain, however, as a low $\rho$ requires a high carrier density (large Fermi surface), while a large $S$  will be associated with low carrier density.

Kuroki and Arita demonstrated theoretically that a large $S$ can be attained in metals having a peculiarly shaped band structure of the so-called ``pudding mold'' type,\cite{Kuroki} which consists of both a dispersive and a flat region.
When the chemical potential of the metal is located within the dispersive region but is close to the flat region, a large asymmetry appears in the velocities of carriers around the chemical potential, producing a large $S$ even in metals with large Fermi surfaces.
Angle-resolved photoemission spectroscopy revealed the existence of a "pudding mold"  band in the thermoelectric oxide Na$_x$CoO$_2$.\cite{Yang,Yang2,Arakane}
This substance simultaneously exhibits large $S$ ($\simeq$ 100 $\mu$V/K) and low $\rho$ ($\simeq$ 200 $\mu\Omega$cm) at 300 K, \cite{Terasaki,Motohashi} resulting in a large PF of $\simeq$ 50 $\mu$W/cmK$^2$, which is comparable to that of the typical thermoelectric material Bi$_2$Te$_3$ (PF $\simeq$ 40 $\mu$W/cmK$^2$).\cite{Caillat}

More recently, Mori {\it et al.} proposed another peculiarly shaped band structure, referred to as the ``corrugated flat band" type, that can attain large values of $S$ in a metallic state.\cite{Mori}
This structure exhibits band dispersion with an overall flatness that extends over the entire Brillouin zone while expressing a series of localized rises and dips that create multiple Fermi pockets (with large carrier density) and a large degree of carrier velocity asymmetry around the chemical potential.
This model successfully explains the enhanced thermoelectric properties of Ir-doped PtSb$_2$ with a cubic pyrite structure (space group $Pa\bar{3}$), which exhibits a maximum power factor of 43 $\mu$W/cmK$^2$ at 400 K.\cite{Nishikubo}
Importantly, Mori {\it et al.} \cite{Mori} predicted that the power factor would continue to increase if the band gap of PtSb$_2$ could be increased, as thermally activated carriers beyond the gap serve to significantly degrade $S$ at high temperatures.
This degradation factor is seen experimentally in PtSb$_2$, in which the value of $S$ starts to decrease with increasing temperature above 400 K owing to the narrow gap (0.06$-$0.08 eV).\cite{Nishikubo}

In this Letter, we report on the enhanced thermoelectric properties of Rh-doped PtAs$_2$ at temperatures above 400 K.
PtAs$_2$ is isotypic and isovalent to PtSb$_2$ but has a larger band gap (0.8 eV).
As in doped PtSb$_2$,\cite{Nishikubo} a large Seebeck coefficient has been observed in the metallic state of doped Pt$_{1-x}$Rh$_x$As$_2$; unlike doped PtSb$_2$, however, the Seebeck coefficient of doped PtAs$_2$ continues to increase with increasing temperature up to 600 K, resulting in  large values of PF (e.g., 65 $\mu$W/cmK$^2$ at 440 K) over this range.

\begin{figure}[t]
\begin{center}
\includegraphics[width=7.5cm]{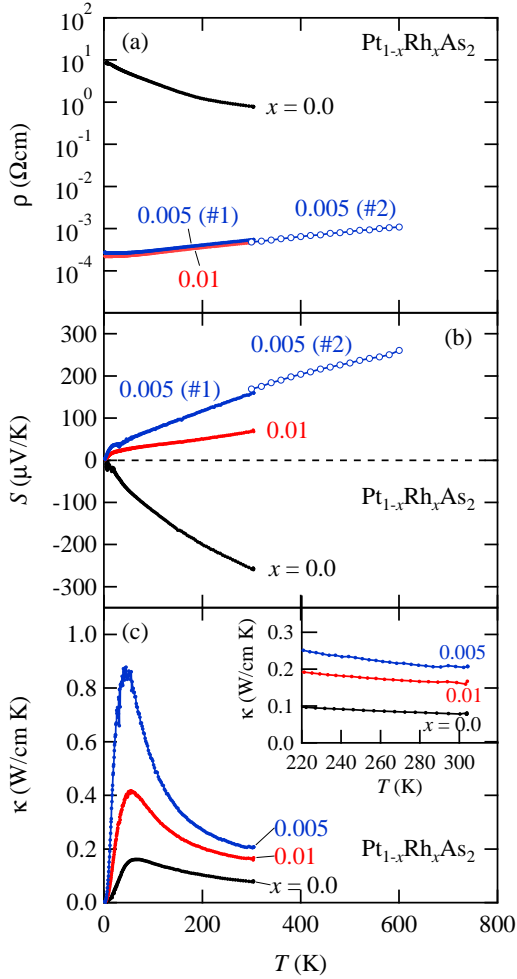}
\caption{\label{fig1}
(Color online) Temperature dependence of (a) electrical resistivity $\rho(T)$, (b) Seebeck coefficient $S(T)$, and (c) thermal conductivity $\kappa(T)$ of polycrystalline Pt$_{1-x}$Rh$_x$As$_2$ with $x$ = 0.0, 0.005, and 0.01.
The inset of (c) shows $\kappa(T)$ data between 220 and 300 K.
}
\end{center}
\end{figure}
\begin{figure}[t]
\begin{center}
\includegraphics[width=7.5cm]{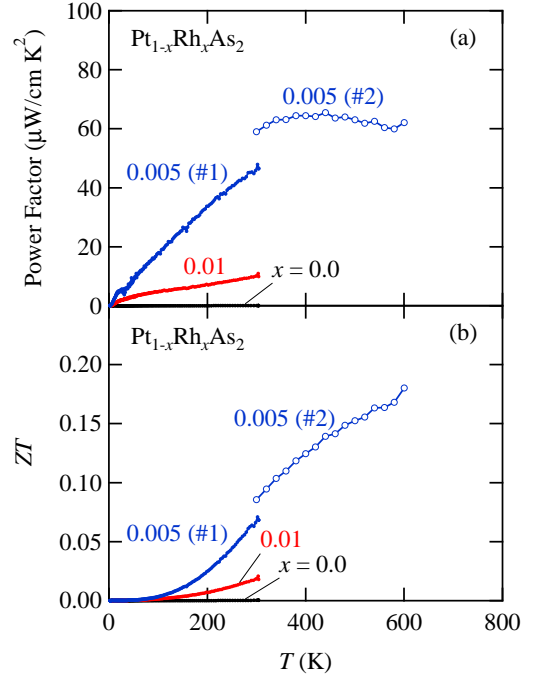}
\caption{\label{fig2}
(Color online) Temperature dependence of (a) power factor (PF) and (b) thermoelectric dimensionless figure of merit $ZT$ of polycrystalline Pt$_{1-x}$Rh$_x$As$_2$ with $x$ = 0.0, 0.005, and 0.01.
We estimated $ZT$ values at $T$ $>$ 300 K using a value of $\kappa$ =  0.207 W/cmK at 300 K.
}
\end{center}
\end{figure}
%
Polycrystalline samples of Pt$_{1-x}$Rh$_x$As$_2$ with $x$ = 0.0, 0.005, and 0.01 were synthesized by means of a solid-state reaction in two steps.
First, stoichiometric amounts of the starting materials Pt (99.99\%), Rh (99.98\%), and As (99.9999\%) were mixed,  ground, and then heated in an evacuated quartz tube at 500$^\circ$C for 10 h and at 700 $^\circ$C for 40 h.
The product was powdered, pressed into pellets, and sintered at 1100$^\circ$C for 10 h.
The resulting samples were characterized by means of powder X-ray diffraction (XRD), and it was confirmed that they consisted of single-phase Pt$_{1-x}$Rh$_x$As$_2$. 
Thermoelectric properties, namely, electrical resistivity $\rho$, Seebeck coefficient $S$, and thermal conductivity $\kappa$ were measured over the temperature range from 2 to 300 K using a physical property measurement system (PPMS, Quantum Design). 
High-temperature values of $S$ and $\rho$ were measured over the temperature range from 300 to 600 K by using, respectively, a commercial thermopower measurement apparatus (MMR Technologies) and  a home-made system with DC four-probe method.

Figure~1(a) shows the temperature dependence of electrical resistivity $\rho$ for Pt$_{1-x}$Rh$_x$As$_2$.
The pristine form of PtAs$_2$ exhibits a semiconducting behavior that is consistent with previous reports,\cite{Johnston} while Rh doping causes both the magnitude and temperature dependence of $\rho$ to change abruptly from semiconducting to metallic.
Values of $\rho$ on the order of 100 $\mu\Omega$cm at 300 K and the positive temperature coefficient of resistivity for Pt$_{1-x}$Rh$_x$As$_2$ ($x$ = 0.005 and 0.01) suggest that these samples have attained a metallic state with the metallic temperature dependence of $\rho$ remaining up to 600 K.
Figure~1(b) shows that the Seebeck coefficient $S$ is negative for $x$ = 0.0, indicating that the majority of the charge carriers in this sample are electrons.
In the non-doped PtAs$_2$, the absolute value of $S$ decreases with decreasing temperature, although this temperature dependence is altered abruptly by Rh doping.
For $x$ = 0.005 and 0.01, the Seebeck coefficient $S$ exhibits a positive value (Fig.~1(b)), which indicates that the majority of charge carriers are holes.
It can be seen that, for $x =$ 0.005, $S$ increases with increasing temperature up to 600 K, at which point it reaches +260 $\mu$V/K; this contrasts strongly with the behavior of Pt$_{1-x}$Ir$_x$Sb$_2$, in which $S$ decreases with increasing temperature above 400 K.\cite{Nishikubo} 
The difference might be attributable to the differential in energy gap magnitudes predicted by Mori {\it et al.},\cite{Mori} which would allow metallic values of $\rho$ to coexist with a large $S$ in Pt$_{1-x}$Rh$_x$As$_2$.

The observed continuous increase in $S$ with increasing temperature results in high values of PF at high temperatures, as can be seen in
Fig.~2(a), which shows the temperature dependence of PF.
Of the samples, $x$ = 0.005 exhibits the highest PF of 65 $\mu$W/cmK$^2$ at approximately 440 K, a value that is higher than those of Pt$_{1-x}$Ir$_x$Sb$_2$ (43 $\mu$W/cmK$^2$)\cite{Nishikubo} and Bi$_2$Te$_3$ (40 $\mu$W/cmK$^2$).\cite{Caillat}
This high PF value is maintained up to 600 K, in marked contrast to Pt$_{1-x}$Ir$_x$Sb$_2$, in which PF decreases significantly above 400 K.\cite{Nishikubo}
It is therefore apparent that a drastic improvement in the high temperature thermoelectric properties of Pt-based pyrites can be attained by replacing Sb with As.

Finally, we examined the thermoelectric efficiency of the doped Pt$_{1-x}$Rh$_x$As$_2$ sample.
Figure~2(b) shows the temperature dependence of the dimensionless figure of merit $ZT$.
Reflecting the high value of $\rho$ shown in Fig.~1(a), $ZT$ is negligibly small at $x =$ 0; as $x$ increases, $ZT$ is enhanced until $x$ = 0.005.
At $x$ = 0.005, $ZT$ increases with increasing temperature until it reaches a maximum value of 0.18 at 600 K.
This high-temperature behavior is drastically enhanced relative to Pt$_{1-x}$Ir$_x$Sb$_2$, in which $ZT$ shows a broad maximum (0.17) at 480 K.\cite{Nishikubo}
In this assessment, we assumed that $\kappa$ is independent of temperature at $T$ $>$ 300 K and used a value of $\kappa$ = 0.207 W/cmK at 300 K to estimate $ZT$ in this upper range.
As shown in Fig.~1(c), the values of $\kappa$ is increased slightly by Rh doping.
The increased value of $\kappa$ = 0.207 W/cmK for $x$ = 0.005, which is approximately one order of magnitude larger than that for Bi$_2$Te$_3$, \cite{Caillat}
is the primary reason that the dimensionless figure of merit ($ZT = S^2T/\rho\kappa$) is suppressed in this result.

Because larger values of PF can be obtained at high temperatures, it is meaningful to reduce the thermal conductivity $\kappa$ in order to enhance $ZT$.
From the measured values of $\rho$, we estimated the electronic thermal conductivity values ($\kappa_e$) using the Wiedemann$-$Franz relation ($\kappa_e = L_0T/\rho$, where $L_0$ is the Lorenz number 2.44 $\times$ 10$^{-8}$ W$\Omega$/K$^2$) and found that, for Pt$_{1-x}$Rh$_x$As$_2$ ($x$ = 0.005), $\kappa_e$ = 0.013 W/cmK at 300 K, with a resultant lattice component, $\kappa_{\rm{lattice}}$, of 0.194 W/cmK.
If this value can be further reduced, the thermoelectric properties of Rh-doped PtAs$_2$ could be improved.
PtAs$_2$ tolerates a variety of chemical substitutions, and the replacement of the As$_2$ molecule, for instance, with the hetero-nuclear diatomic GeSe molecule, which is isovalent to As$_2$, might be useful in suppressing $\kappa_{\rm{lattice}}$. 
If $ZT$ can be enhanced, the present compound could be applied to thin-film thermoelectric devices, though the use of Pt and Rh is not compatible with the applications in a large scale.

In this study, polycrystalline samples of Pt$_{1-x}$Rh$_x$As$_2$ with $x$ = 0.0, 0.005, and 0.01 were prepared and their thermoelectric properties were investigated at 2$-$600 K.
The doped samples exhibited metallic conductivity and a large Seebeck coefficient, which combined to enhance the power factor to approximately 65 $\mu$W/cmK$^2$ at 440 K for Pt$_{1-x}$Rh$_x$As$_2$ with $x$ = 0.005.
As predicted by Mori {\it et al.}, materials that combine a ``corrugated flat" band structure with a wider band gap have high thermoelectric properties.
Such a peculiarly shaped band structure has been predicted in PtAs$_2$ by the first principles
calculation.\cite{Mori2}

The authors are grateful to H. Usui and K. Kuroki for valuable discussion and to K. Kajisa and K. Niitani for their help in the high-temperature measurements.
Part of this work was performed at the Advanced Science Research Center, Okayama University.
This work was partially supported by a Grant-in-Aid for Scientific Research (C) (25400372) from the Japan Society of the Promotion of Science (JSPS) and the Funding Program for World-Leading Innovative R\&D on Science and Technology (FIRST Program) from JSPS.

\end{document}